\pdfoutput=1
\documentclass[conference,a4paper]{IEEEtran}
\usepackage{cite}
\usepackage[hyphens]{url}
\usepackage[hidelinks]{hyperref}
\hypersetup{breaklinks=true}

\usepackage{comment}
\usepackage{graphicx}
\usepackage{verbatim}

\usepackage{siunitx}

\usepackage{tikz}
\usepackage{tikzscale}
\usepackage{pgfplots}
\usepackage{float}
\usepackage{todonotes}

\usepackage[labelfont=bf]{caption}

\usepackage{subcaption}

    \usepackage{pgfplots}
    \usepgfplotslibrary{groupplots}
    \pgfplotsset{compat=1.11,
    /pgfplots/ybar legend/.style={
    /pgfplots/legend image code/.code={%
       \draw[##1,/tikz/.cd,yshift=-0.25em]
        (0cm,0cm) rectangle (3pt,0.8em);},
   },
}

\usepackage{amssymb}
\usepackage{pifont}
\usetikzlibrary{patterns,arrows}
\usepackage{color}

\usepackage{subfloat}

\makeatletter
\IEEEtriggercmd{\reset@font\normalfont\fontsize{7.95pt}{8.0pt}\selectfont}
\makeatother
\IEEEtriggeratref{1}

\newcommand{\ballnumber}[1]{\tikz[baseline=(myanchor.base)] \node[circle,fill=.,inner sep=1pt] (myanchor) {\color{-.}\bfseries\footnotesize #1};}
\ifCLASSINFOpdf
\else
\fi
\begin{document}

\newpage

%
\title{NMPO: Near-Memory Computing Profiling and Offloading}

\author{

Stefano Corda$^{1,2}$/Madhurya Kumaraswamy$^{1}$, Ahsan Javed Awan$^3$, Roel Jordans$^1$, Akash Kumar$^2$, Henk Corporaal$^1$\\

\vspace{-0.4cm} \normalsize $^1$Eindhoven University of Technology $^2$Technische Universität Dresden $^3$Ericsson Research \\\\

\{s.corda, r.jordans, h.corporaal\}@tue.nl,
m.kumaraswamy@student.tue.nl\\
stefano.corda@mailbox.tu-dresden.de
akash.kumar@tu-dresden.de
ahsan.javed.awan@ericsson.com
}

\maketitle

\thispagestyle{plain}
\pagestyle{plain}

\definecolor{dgreen}{rgb}{0.00, 0.75, 0.00}
\definecolor{dred}{rgb}{0.75, 0.00, 0.00}
\definecolor{dblue}{rgb}{0.00, 0.00, 0.75}
\newcommand{\stefano}[1]{[{\color{blue}Stefano: #1}]}
\newcommand{\madhurya}[1]{[{\color{orange}Madhurya: #1}]}


\begin{abstract}
Real-world applications are now processing big-data sets, often bottlenecked by the data movement between the compute units and the main memory. Near-memory computing (NMC), a modern data-centric computational paradigm, can alleviate these bottlenecks, thereby improving the performance of applications. The lack of NMC system availability makes simulators the primary evaluation tool for performance estimation. However, simulators are usually time-consuming, and methods that can reduce this overhead would accelerate the early-stage design process of NMC systems. This work proposes Near-Memory computing Profiling and Offloading (NMPO), a high-level framework capable of predicting NMC offloading suitability employing an ensemble machine learning model. NMPO predicts NMC suitability with an accuracy of 85.6\% and, compared to prior works, can reduce the prediction time by using hardware-dependent applications features by up to 3 order of magnitude.

\end{abstract}


%
\IEEEpeerreviewmaketitle

\section{Introduction}
\label{sec:introduction}

Modern big-data applications comprise machine learning, radio-astronomical imaging, and bioinformatics algorithms \cite{7310708}. These workloads usually impose high compute and data requirements, which may cause bottlenecks. Most of the applications that process large datasets frequently stall in the cache hierarchy due to the data movement between the main memory, and the processor \cite{awan2017performance}.
A proposed solution to this problem is \emph{near-memory computing} (NMC) \cite{8491877,SINGH2019102868,10.1145/3299874.3322805}, which is an opposite computing paradigm to the classical compute-centric being data-centric and performs the computation near the memory, avoiding the data-movement mentioned above. NMC is possible by the recent advancement in memory technologies. Indeed, technologies such as 3D-stacked memory \cite{3Dstack_tech} have higher bandwidth, a large number of channels, reduced power consumption, and the possibility to place accelerators on the logic layer of the memory itself \cite{HMC,HBM}.
Prior works show how NMC can efficiently be employed to improve the performance of applications such as graph processing \cite{7920847}, numerical simulations \cite{8715088}, machine learning \cite{9138955}, image processing \cite{9138985}, and radio-astronomical imaging \cite{9134089}.

Nevertheless, due to the poor availability of NMC systems, mainly consisting of prototypes, it is challenging to profile and evaluate the suitability of NMC architectures. Predominantly, system designers use simulation techniques to evaluate workloads \cite{10.1109/ISCA45697.2020.00072}. These simulators need to be configured for each new DRAM technology and are time-consuming: they can take up to days or even weeks for real-world application with ever-growing big-data datasets. 
A systematic methodology for identifying any application's NMC suitability helps the programmer in faster early design stage exploration. 
Therefore, this work proposes a high-level Near-Memory computing Profiling and Offloading (NMPO) framework for evaluating the NMC suitability of applications by employing an ensemble machine learning algorithm. 
NMPO's goal is to provide a quick estimation of the NMC suitability by training a Random Forest (RF) model with micro-architecture dependent profiling characteristics.
NMPO trains and predicts using hardware-dependent characteristics that are usually faster by approximately 2 to 3 orders of magnitude \cite{10.1145/2742854.2742859} compared to the platform-independent features employed in related work \cite{8806888}. This huge overhead difference is due to the large memory requirements that hardware-independent analysis needs for certain analysis, while hardware-dependent characterization relies on fast hardware performance counters. Despite this benefit, NMPO still needs to run the time-consuming NMC simulations for training the ML model, and it also needs information about the NMC performance.

Summarizing the paper's contributions:
\begin{itemize}
    \item NMPO, a fast high-level profiling and offloading framework for NMC systems, to analyze an application in the early design phases and evaluate if it is suitable for NMC offloading. We employ hardware-dependent profiling techniques and ensemble machine learning models with feature selection and hyper-tuning to build the framework.
    \item NMPO predicts NMC suitability with an accuracy of 85.6\%, and it reduces the prediction time by 2 to 3 order of magnitude compared to the state-of-the-art NMC performance prediction model \cite{8806888}.
\end{itemize}

The paper is structured as follows: \emph{Section \ref{sec:background}} presents the essential concepts on application characterization, NMC simulation and machine learning models. In \emph{Section \ref{sec:methodology}} we explain the adopted methodology. Then, \emph{Section \ref{sec:results}} shows our framework evaluation results in terms of accuracy and speed. Related works are discussed in \emph{Section \ref{sec:relatedwork}} and \emph{Section \ref{sec:conclusion}} concludes the paper.

\section{Background}
\label{sec:background}

This section reports the necessary background about performance monitoring counters (\emph{\ref{subsec:applicationcharacterization}}), NMC simulation (\emph{\ref{subsec:nmcsimulation}}) and ensemble machine learning models (\emph{\ref{subsec:ensembleml}}).

\subsection{Application characterization}
\label{subsec:applicationcharacterization}

Key application features that are used later for taking offloading decisions can be collected in different ways. The quicker and easier way of evaluating an application on a traditional CPU is using hardware performance monitoring units (PMUs). Modern CPUs have specific programmable components programmed to gather information from different locations of the chip. Currently, a wide range of tools and libraries can be employed for this task, such as PAPI \cite{PAPI}, LIKWID \cite{LIKWID}, and perf \cite{perf}. Perf is a ready-to-use utility available in most current Linux distributions. This utility collects an enormous amount of information from the analyzed application, such as cache misses, Clock cycles per Instructions (CPI), and floating-point operations. We summarize the main features that are collected in \emph{Table \ref{table:perfmetrics}}.

\begin{table}[h]
\centering
\captionof{table}{\emph{Perf event list of the Host Machine.}}
\scalebox{0.95}{
\begin{tabular}{|l|l|l|l|}
\hline
\textbf{Event name}                           & \textbf{Units}            &\textbf{Event name}                           & \textbf{Units}                                                                                                              \\ \hline
power/energy-pkg/                    &     Joules             &       L1-dcache-loads                      & countof                                                                                               \\ \hline
power/energy-psys/                   &     Joules             &      L1-dcache-stores                     & countof                                                                                                  \\ \hline
power/energy-ram/                    &      Joules          &      L1-icache-load-misses                & countof                                                                                                     \\ \hline
uncore\_imc/data\_reads/             &     MiB             &                                                                  LLC-load-misses                      & countof                                      \\ \hline
uncore\_imc/data\_writes/            &      MiB            &  context switch & countof                                                                                                    \\ \hline
fp\_arith\_inst\_retired             & Gflops   & App execution time & seconds \\ \hline
                                
branch-instruction/branches & countof &    LLC-loads                            & countof                                                                                                   \\ \hline
branch-misses                        & countof                 &           LLC-store-misses                     & countof                                                                                              \\ \hline
cache-misses                         & countof                 &    LLC-stores                           & countof                                                                                                         \\ \hline
cpu-cycles OR cycles                 & countof                 &      branch-load-misses                   & countof                                                                                                    \\ \hline
instructions                         & countof                 &    branch-loads                         & countof                                                                                                     \\ \hline
L1-dcache-load-misses                & countof                 &   Instructions/cycle  &  IPC                                                                                            \\ \hline

\end{tabular}
}

\label{table:perfmetrics}
\end{table}

\subsection{NMC simulation}
\label{subsec:nmcsimulation}
\begin{figure}[h]
    \centering
    \includegraphics[width=9cm]{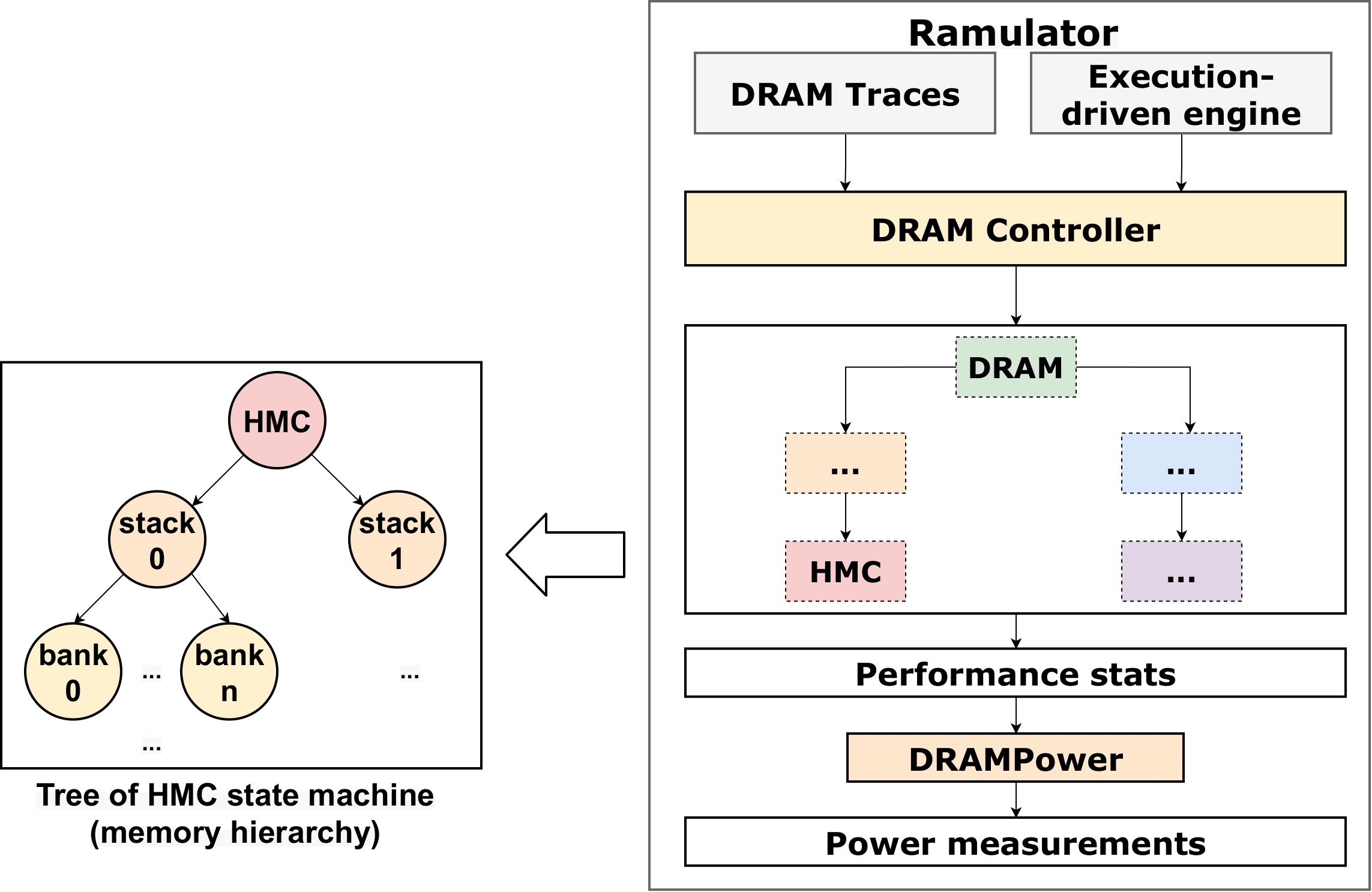}
    \caption{\emph{High-level overview of Ramulator.}}
    \label{fig:ramulator}
    \end{figure}

Since NMC systems adoption is still not widespread, simulators are necessarily employed to determine their performance. Extended versions of Ramulator \cite{8806888,9138955,9138985,10.1109/ISCA45697.2020.00072} are utilized because of its easy extendibility, speed and accuracy. Ramulator is a cycle-accurate and portable memory simulator that simulates a wide range of modern DRAM technologies such as HBM (High Bandwidth Memory), HMC (Hyper Memory Cube), and WideIO.
\emph{Fig. \ref{fig:ramulator}} shows a high-level representation of Ramulator. It consists of a memory controller that takes the simulation's input. This input can be a set of memory traces generated by a CPU simulator such as Zsim \cite{10.1145/2485922.2485963}, which is called standalone mode, or it can be generated by an execution-driven engine such as Gem5 \cite{10.1145/2024716.2024718}, which is named integrated mode. Ramulator's core consists of a tree of DRAM state-machines (left side of \emph{Fig. \ref{fig:ramulator}}), where each node is a class instance such as HMC that derives its properties from its parents' nodes. Each DRAM class has a hierarchy of banks, channels, ranks, etc., representing different nodes having a specific label as property.
Ramulator-PIM, an extended version of Ramulator, can simulate computing units such as Out of Order (Ooo) cores on the logic layer of 3D-stacked memory.

For the evaluation of power consumption metrics Ramulator is integrated with DRAM power models such as DRAMPower \cite{DRAMPower}. In \emph{Table \ref{table:ramulatormetrics}} we summarize the main performance metrics that can be extracted using Ramulator-PIM.

\begin{table}[h]

\centering
\captionof{table}{\emph{NMC system performance metrics.}}
\label{table:ramulatormetrics}
\begin{tabular}{|l|l|l|}
\hline
\textbf{Statistic}                  & \textbf{Units}            & \textbf{Category}      \\ \hline
ramulator.cpu\_cycles       & cycle             & Ramulator-PIM \\ \hline
ramulator.ipc               & Instruction/cycle & Ramulator-PIM \\ \hline
ramulator.cpu\_instructions & countof             & Ramulator-PIM \\ \hline
ramulator.total\_time       & ns                & Ramulator-PIM \\ \hline
Average Power               & mW                & DRAM Power    \\ \hline
Total Trace Energy          & pJ                & DRAM Power    \\ \hline
\end{tabular}

\end{table}

\subsection{Ensemble machine learning}
\label{subsec:ensembleml}

Complex decisions such as application offloading to suitable accelerators may require sophisticated tools such as machine learning (ML) prediction models. These models are usually trained on a section of the available features dataset and tested on the remaining part. Then, they are employed to predict, make decisions or classify an unknown dataset. While simple models such as a Decision Tree can be effective, in the case of many features, ensemble ML models are more accurate \cite{10.5555/648054.743935}. Ensemble ML models consist of several simple models trained on a different and random subset of the training dataset. The final decision, classification, or prediction is then made by evaluating all the simple models' results by selecting the most common outcome. 

Random Forest (RF) \cite{10.1023/A:1010933404324} is an ensemble ML model that consists of a set of decision trees. RF uses either a categorical response variable, referred to in \cite{Scikit} as ``classification", or a continuous response referred to as ``regression". Similarly, the predictor variables can be either categorical or continuous. The decision trees are partitioned based on binary recursion. The predictor space uses a sequence of binary splits to partition. The root node contains the whole list of predictors. The splitting criterion gives a measure of ``goodness of fit" (regression) or ``purity" (classification) for a node, with large values representing poor fit (regression) or an impure node (classification).

RF model performance is boosted by tuning the hyper-parameters, which are characteristics of the model that can impact model accuracy and computational efficiency. These values are set before fitting the model and optimized through trial and error methods like grid search and random search. Multiple models are fitted with several hyper-parameter value sets, their performances are compared, and the best performing one is chosen. The popular hyper-parameters tuned for Random Forest models are; the number of decision trees (N\_estimators), the number of features to resample  (Max\_features), the depth of each tree in the forest (Max\_depth), the minimum number of samples required to split each node (Min\_samples\_split) and the minimum number of samples required for each leaf (Min\_samples\_leaf).

\begin{figure*}[]
    \centering
    \includegraphics[width=18cm]{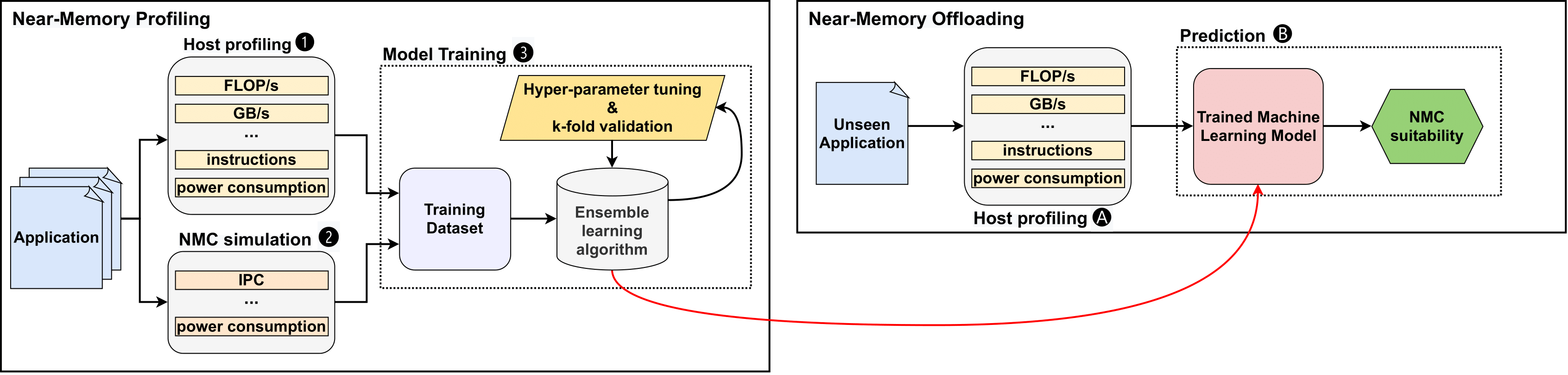}
   \caption{\emph{Near-Memory Computing Profiling and Offloading (NMPO) overview.}
    \label{fig:NMPO}}
\end{figure*}

\begin{table*}[]
\centering
\caption{\emph{Applications and parameters.} \label{tab:application_and_parameters}}
\scalebox{0.95}{
\begin{tabular}{ccc|cccccccc|cc}
\hline
\multicolumn{3}{c}{\textbf{Application}} & \multicolumn{8}{c}{\textbf{Datasets levels}}&
\multicolumn{2}{c}{\textbf{Time (min)}}\\
\hline
Name & Task & Threads & 1 & 2 & 3 & 4 & 5 & 6 & 7 & \textbf{Test}& ML &RT\\
\hline
atax  & Computes A\^T times Ax & 8,16 & 4000 & 6000 & 8000 & 10000 & 12000 & 14000 & 16000  & 17000 & 3.25&180 \\
chol  & Decomposes a matrix to triangular matrices & 8,16 & 1024 & 1500  & 2000 & 2200 & 2600  & 3000 & 3400 & 4000 & 6&720\\
doit  &  Multiresolution ADaptive NumErical Scientific & 8,16 & 75 & 100  & 128 & 150 & 200 & 256 & 300  & 350 & 6.25&5760\\
gemv  & Multiple matrix-vector multiplication &8,16 & 4000  & 6000  & 8000 &  10000 & 12000 & 14000  &  16000 & 18000 & 7.15&186\\
gesu  & Summed matrix-vector multiplications&8,16 & 4000  & 6000  & 8000 &  10000 & 12000 & 14000  &  16000 & 18000 & 8.35&202\\
mvt  & Matrix Vector
Product & 8,16 & 4000  & 6000  & 8000 &  10000 & 12000 & 14000  &  16000 & 18000 & 7.56&173\\
syrk  &Symmetric rank k update & 8,16& 1024 & 1500  & 2000 &  2500 & 2750 & 3000 & 3500 & 4000 & 9.32&4568 \\
syr2k & Symmetric rank 2k update& 8,16& 1024 & 1500  & 2000 &   2500 & 2750 & 3000 & 3500 & 4000 & 8.4&4898 \\
trmm  &Triangular matrix multiplication & 8,16 & 1024  & 1500 & 2000 & 2500 & 2750 & 3000 & 3500 & 4000 & 7.35&5280 \\
\hline
grid  & Radio-astronomical visibilities gridder & 8,16& 128  & 256 & 512 & 2048 & 2560 & 3072 & 3584 & 4096 & 8.15 & / \\
degrid & Radio-astronomical visibilities degriddeg & 8,16& 128  & 256 & 512 & 2048 & 2560 & 3072 & 3584 & 4096 & 8.32 & / \\
\hline
\end{tabular}
}
\end{table*}

\section{Methodology}
\label{sec:methodology}

The NMPO framework and the experimental setup are described respectively in \emph{Subsection \ref{subsec:nmpo}} and in \emph{Subsection \ref{subsec:experimentalsetup}}.

\subsection{NMPO}
\label{subsec:nmpo}

NMPO (see \emph{Fig. \ref{fig:NMPO}}) consists mainly of two separate parts: the first one for characterizing the application and training the machine learning model and the second one where the offloading decision is taken by employing the ML model's prediction result.
More precisely, in the first phase, the applications are characterized on the host system (\ballnumber{1}) employing PMUs and collecting information as reported in \emph{Table \ref{table:perfmetrics}}. Then, the applications are simulated on the NMC system (\ballnumber{2}), using Ramulator and DRAMPower to obtain the performance measurements (see \emph{Table \ref{table:ramulatormetrics}}). The performance metrics gathered from these steps are applied to evaluate the NMC offloading suitability. Thus, we label the data for the machine learning model by our criteria of offloading based on Energy-Delay-Product speedup, which is computed as follows:
\begin{equation}
\text{EDP\_speedup = Host\_EDP/ NMC\_EDP}
\end{equation}
 Accordingly, for the collected training data, we label the offloading decision as ``yes" if EDP\_speedup $>$ 2, ``maybe" if 1 $<$ EDP\_speedup $>$ 2 and ``no" if EDP\_speedup $<$ 1.
Finally, the machine learning model is trained (\ballnumber{3}) using the previous analysis metrics. We employ k-fold validation to evaluate the ML model. For each of the K folds, the model is trained on the remaining (K - 1) folds, which are considered training data and tested on the remaining data or the left-out fold, which serves as the testing data. The performance of the machine learning model is evaluated as the average performance over K-iterations of cross-validation. 
The hyper-parameters, which are the ML algorithm variables, are tuned to optimize the prediction model's accuracy. 
The application offloading of the unseen application is performed by first profiling the application \ballnumber{A}, similarly to \ballnumber{1}, on the host system with PMUs (see \emph{Table \ref{table:perfmetrics}}). Then, the trained ML model uses the extracted features to predict the offloading decision on an NMC system \ballnumber{B}.
More precisely, since the key feature is the Ramulator IPC (see \emph{Fig. \ref{fig:Correlation plot}}), the ML model predicts this key feature for unseen application by employing RF regression model by using only the host system characterization and later predicts the NMC suitability by classifying the results.
The performance of the machine learning model evaluates as the average performance over K-iterations of cross-validation. For example, let the RF ensemble model compute the regression error in predicting the $k$th part using RMSE and cross-validation score (CV) as:  

$$ RMSE_{k}=\sqrt{\frac{\sum_{i \in k t h \text { part}}\left(\text {Predicted}_{i}-\text {Actual}_{i}\right)^{2}}{N}} $$
$$ C V=\frac{1}{K} \sum_{k=1}^{K} RMSE_{k} $$

We shaped the NMC offloading decision as a classification problem, where the key error metric is accuracy that correspond to the ratio of correct prediction and the total number of prediction:
     
$$ \text{Accuracy} = \frac{\text{Number} \; \text{of} \; \text{correct} \; \text{predictions}}{\text{Total} \; \text{number} \; \text{of} \; \text{predictions}} $$

We also applied the confusion matrix as an alternative tool to better visualize the same information. In the confusion matrix, each row of the matrix represents the instances in a predicted class, each column represents the instances in an actual class. The confusion matrix is named since it makes it easy to see if the system confuses one class for another.

\subsection{Experimental setup}
\label{subsec:experimentalsetup}

\begin{figure}[H]
    \centering
    \includegraphics[width=9cm]{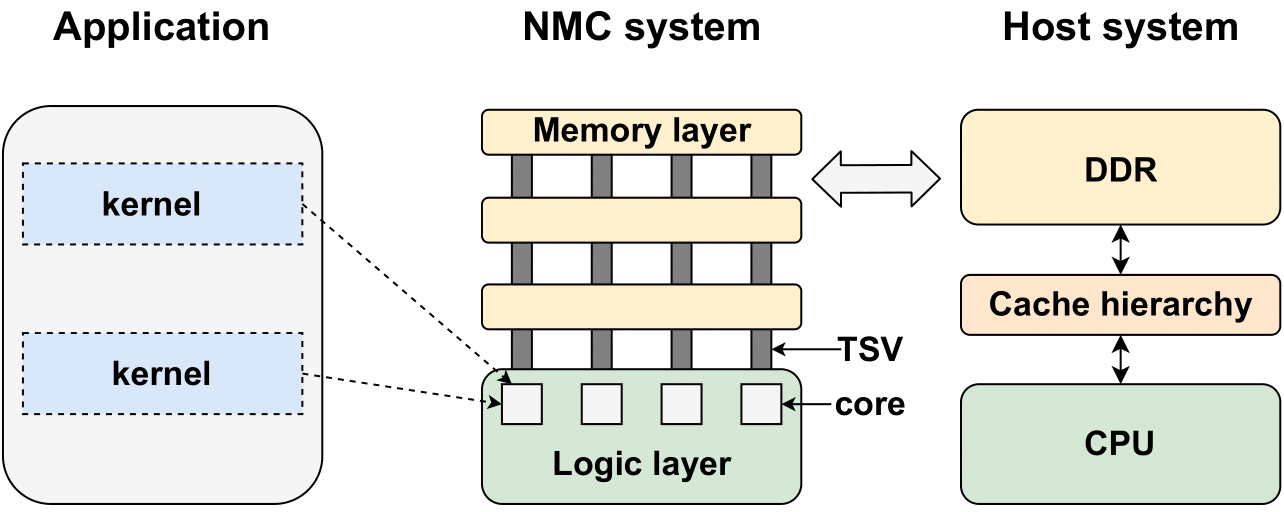}
   \caption{\emph{System overview.}
    \label{fig:system over}}
\end{figure}

Our experimental setups consist mainly of a host processor (see \emph{Fig. \ref{fig:system over}}), an Intel i9 9900K, and an NMC system with out-of-order (Ooo) cores placed on the logic layer of the HMC memory. The details of the host and NMC system are presented in \emph{Table \ref{tab:systemparams}}.

\begin{table}[H]
\centering
\caption{\emph{Systems parameters.} \label{tab:systemparams}}
\scalebox{1}{
\begin{tabular}{l|l}
\multicolumn{2}{c}{\textbf{Host system}} \\
\hline
\textbf{Intel i9} & 8 cores, 2 threads per core, 1 socket, \SI{4.7}{GHz},\\
\textbf{9900K} & \SI{16}{MB} L3 cache, \SI{64}{GB} DDR4 \SI{2666}{MHz},\\
\hline
\multicolumn{2}{c}{\textbf{NMC system}} \\
\hline
\textbf{Ramulator} & 8 single issue Ooo cores \SI{1.25}{GHz}\\
 &  2-way, 2 cache-lines, \SI{64}{B} per cache-line\\
 & 32 vaults, 8 stacked-layers, \SI{256}{B} row buffer, \SI{4}{GB} HMC\\
 & 16-bit full duplex high-speed SerDes I/O link \SI{15}{GBps}\\
\hline
\end{tabular}
}
\end{table}

The applications are profiled on the host system five times, extracting mean values with the perf package available with Ubuntu 18.04.
The NMC system is simulated with Ramulator-PIM \cite{ramulator-pim} once, since the results do not vary in different runs. Power and time parameters for HMC are derived from \cite{DRAMSpec1,DRAMPower} and fed to the NMC simulator.
As a benchmark, we selected a set of application from Polybench since it consists of simple mathematical operations extensively used in modern applications, and are also commonly used in NMC related work \cite{7756764}. We selected implementation of the benchmark using OpenMP \cite{polybench-acc,6339595}, in order to exploit parallelism on the CPU.
Aside from synthetic benchmarks, we use the current state-of-the-art gridding, and degridding algorithm for radio-astronomical imaging Image Domain Gridding (IDG) release 0.7 \cite{astron-idg,VEENBOER2020100386}. 
As shown in \emph{Fig. \ref{fig:system over}}, we analyzed only the kernel of interest. While Polybench applications have just one kernel, IDG contains different kernel such as gridder and degridder.
The benchmarks, their parameters, and the value associated with the different dataset sizes are listed in \emph{Table \ref{tab:application_and_parameters}}. The datasets are carefully chosen to be large enough to generate DRAM accesses and evaluate whether the application is really suitable for NMC. We also reported the time spent by the machine learning (ML) for the training, hyper-tuning and prediction and the Ramulator simulation time for collecting training data (RT).

\section{Experimental Results and evaluation}
\label{sec:results}
In this section, we discuss the results of application profiling and offloading. Further, empirical evidence in terms of validation and error metrics
of the prediction models is presented. Finally, the prediction models are applied to the test cases for identifying the NMC suitability for a target application, thus aiding the users in early design stage explorations.

\subsection{Application profiling}
\label{subsec:applicationprofiling}
This stage provides the training data required to build and test our machine learning model \emph{Section \ref{subsec:nmpo}}. Applications with chosen datasets levels in \emph{Table \ref{tab:application_and_parameters}} are profiled to collect various statistics from perf, Ramulator-PIM and DRAMPower as discussed in \emph{Section \ref {subsec:applicationcharacterization}}. The roofline model \cite{williams2009roofline,ofenbeck2014applying} is a method for capturing the compute-memory ratio of computation and determines if the application is compute-bound or memory bound. The roofline model shows the application’s achieved performance (GFLOP/s) and arithmetic intensity (FLOP/Byte) against the machine’s maximum achievable performance.

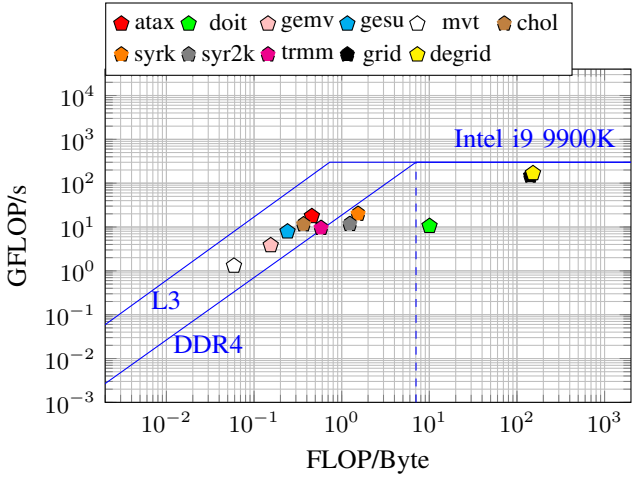
\begin{figure}[h]
\centering
\begin{tikzpicture}
\begin{axis}[width=8.5cm, height=6cm,
legend pos=north west,
legend style={style={font=\fontsize{9}{5}\selectfont}},
legend style ={at={(0,1.2)}},
legend columns=6,
legend style={cells={align=left}},
xmax=2000,
ymin=0.001,
xmin=0.002,
ymax=40000,
ymode=log, 
xmode=log,
ytick distance=10,
ymajorgrids=true,
xmajorgrids=true,
xminorgrids=true,
yminorgrids=true,
xlabel=FLOP/Byte,
ylabel=GFLOP/s
]
\pgfplotsset{every axis/.append style={
                    tick label style={font=\small}  
                    }}

\node[above,blue] at (160, 400) {Intel i9 9900K};

\node[above,blue] at (0.01, 0.08) {L3};
\node[above,blue] at (0.03, 0.008) {DDR4};

\addplot+[only marks,mark=pentagon*,mark options={scale=1.5, fill=red},text mark as node=true, color=black] coordinates {
 (0.457,17.639)
};
\addlegendentry{atax}

\addplot+[only marks,mark=pentagon*,mark options={scale=1.5, fill=green},text mark as node=true, color=black] coordinates {
 ( 10.053,10.527)
};
\addlegendentry{doit}

\addplot+[only marks,mark=pentagon*,mark options={scale=1.5, fill=pink},text mark as node=true, color=black] coordinates {
(0.155,3.828)
};
\addlegendentry{gemv}

\addplot+[only marks,mark=pentagon*,mark options={scale=1.5, fill=cyan},text mark as node=true, color=black] coordinates {
 (0.241,7.871)
};
\addlegendentry{gesu}

\addplot+[only marks,mark=pentagon*,mark options={scale=1.5, fill=white},text mark as node=true, color=black] coordinates {
 ( 0.059,1.302)
};
\addlegendentry{mvt}
\addplot+[only marks,mark=pentagon*,mark options={scale=1.5, fill=brown},text mark as node=true, color=black] coordinates {
 ( 0.367,11.469)
};
\addlegendentry{chol}

\addplot+[only marks,mark=pentagon*,mark options={scale=1.5, fill=orange},text mark as node=true, color=black] coordinates {
 (1.539,19.902)
};
\addlegendentry{syrk}

\addplot+[only marks,mark=pentagon*,mark options={scale=1.5, fill=gray},text mark as node=true, color=black] coordinates {
 (1.239,11.692)
};
\addlegendentry{syr2k}

\addplot+[only marks,mark=pentagon*,mark options={scale=1.5, fill=magenta},text mark as node=true, color=black] coordinates {
 (0.582,9.517)
};
\addlegendentry{trmm}

\addplot+[only marks,mark=pentagon*,mark options={scale=1.5, fill=black},text mark as node=true, color=black] coordinates {
 (142.8021891,147.6472206)
};
\addlegendentry{grid}

\addplot+[only marks,mark=pentagon*,mark options={scale=1.5, fill=yellow},text mark as node=true, color=black] coordinates {
 (152.4472574, 168.2361104)
};
\addlegendentry{degrid}





\addplot[
    color=blue,
    ]
    coordinates {
    (0.001,0.001)(7.05,300.8)(2000,300.8)
    };
    
    \addplot[
    color=blue,
    ]
    coordinates {
    (0.000001,0.000001)(0.73,300.8)(2000,300.8)
    };
    
\addplot[
    color=blue,
    style=dashed,
    ]
    coordinates {
    (7.05,0.001)(7.05,300.8)
    };    
    
\end{axis}
\end{tikzpicture}
\label{fig:roofline}
\caption{\emph{Roofline model of test cases using 16 threads.}}

\label{fig:roofline}
\end{figure}

In \emph{Fig. \ref{fig:roofline}} the roofline model of 16 threads test datasets is plotted as an example. Application such as gridder, degridder and doitgen are compute-bound; Symmetric rank update algorithms (syrk and syr2k) are in the DRAM-bound region, whereas the rest of the applications are L3-cache bounded. This tool helps to demonstrate the heterogeneity of the benchmark employed.

\begin{figure}[h]
    \centering
    \includegraphics[width=8.5cm]{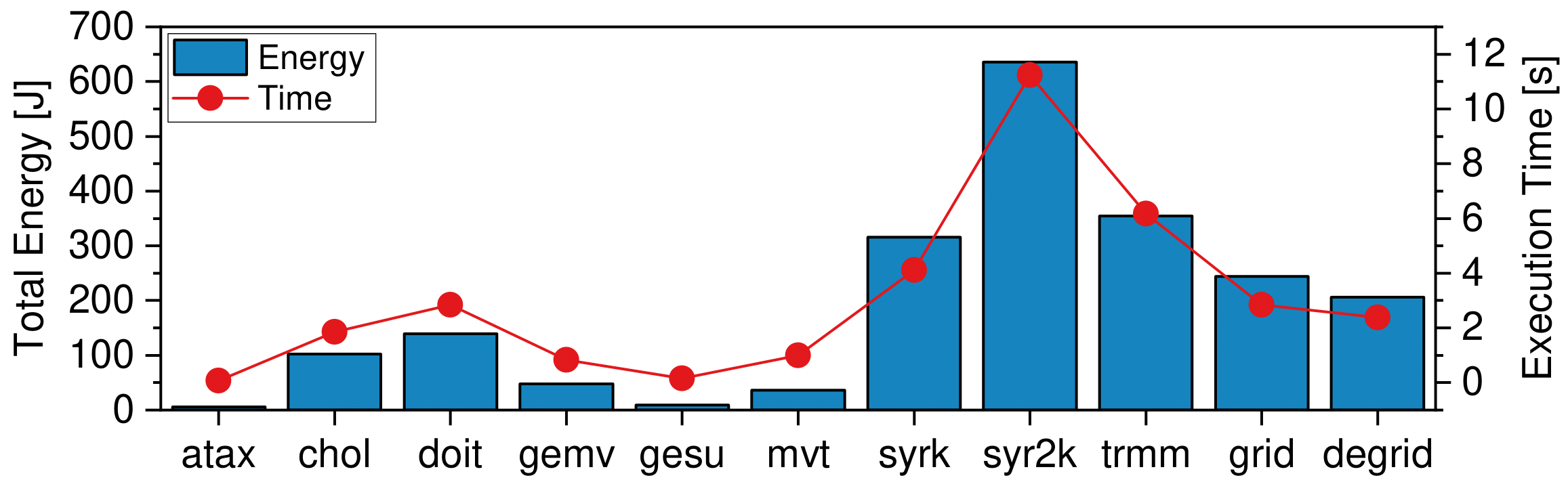}
   \caption{\emph{Execution time and total energy of test cases on Intel i9  using 16 threads.}
    \label{fig:energytime}}
\end{figure}

\begin{figure}[h]
    \centering
    \includegraphics[width=9cm]{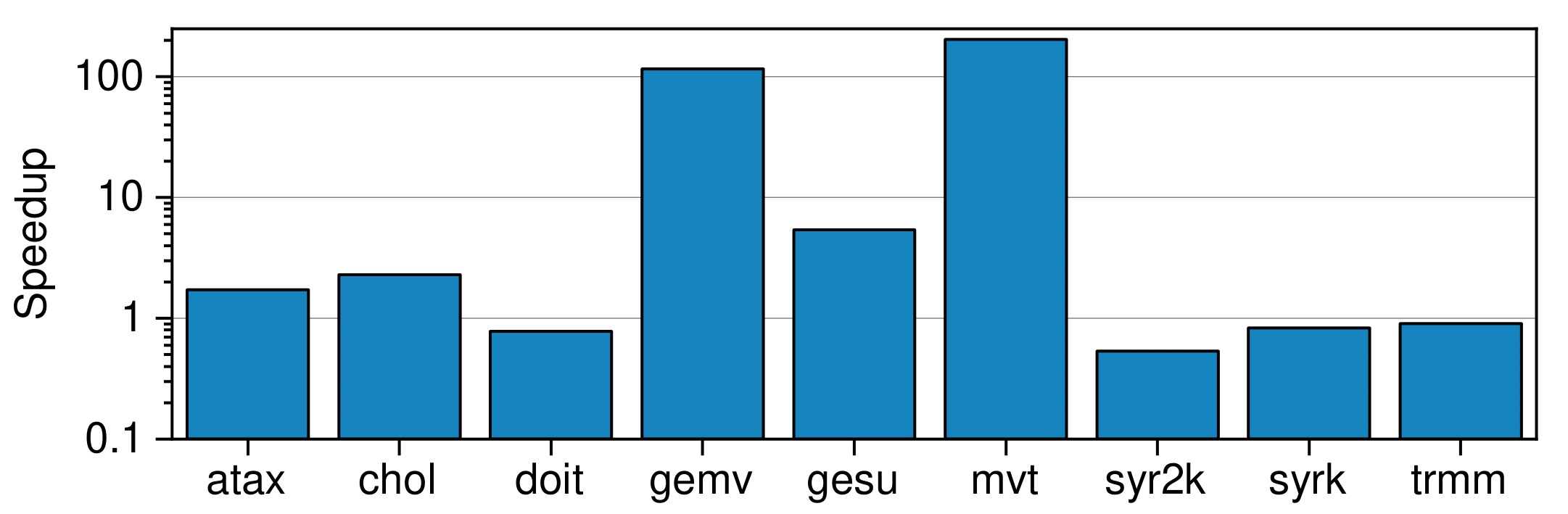}
   \caption{\emph{EDP Speedup of Polybench test cases using 16 threads.}
    \label{fig:edpspeedup}}
\end{figure}

In \emph{Fig. \ref{fig:energytime}} Total energy (J) vs Execution time (s) is depicted for all test cases for 16 threads showing the above-mentioned applications heterogeneity. The proposed work uses the energy-delay product (EDP) of host and NMC, where energy is the total energy consumption of cores and delay is the amount of time for executing applications. Then, we compute the EDP Speedup (\emph{Fig. \ref{fig:edpspeedup}} shows only the EDP speedup for the test cases using 16 threads) for each training dataset.

\subsection{Application offloading}
\label{subsec:offloading}
\subsubsection{Feature selection}
\label{subsec:Validation and EM}
Feature selection methods are intended to reduce the number of input variables to ones that are the most beneficial to a model to predict the target variable. This technique is employed to improve estimators’ accuracy scores or boost their performance on very high-dimensional data sets. In our analysis, we selected the essential features using Pearson correlation. It is represented by a number between -1 and 1 that indicates the extent to which two variables are linearly related. A value closer to 1 implies a stronger positive correlation, and a value closer to -1 indicates a negative correlation. 
    \begin{figure}[h]
    
    \centering
    \includegraphics[width=8.5cm]{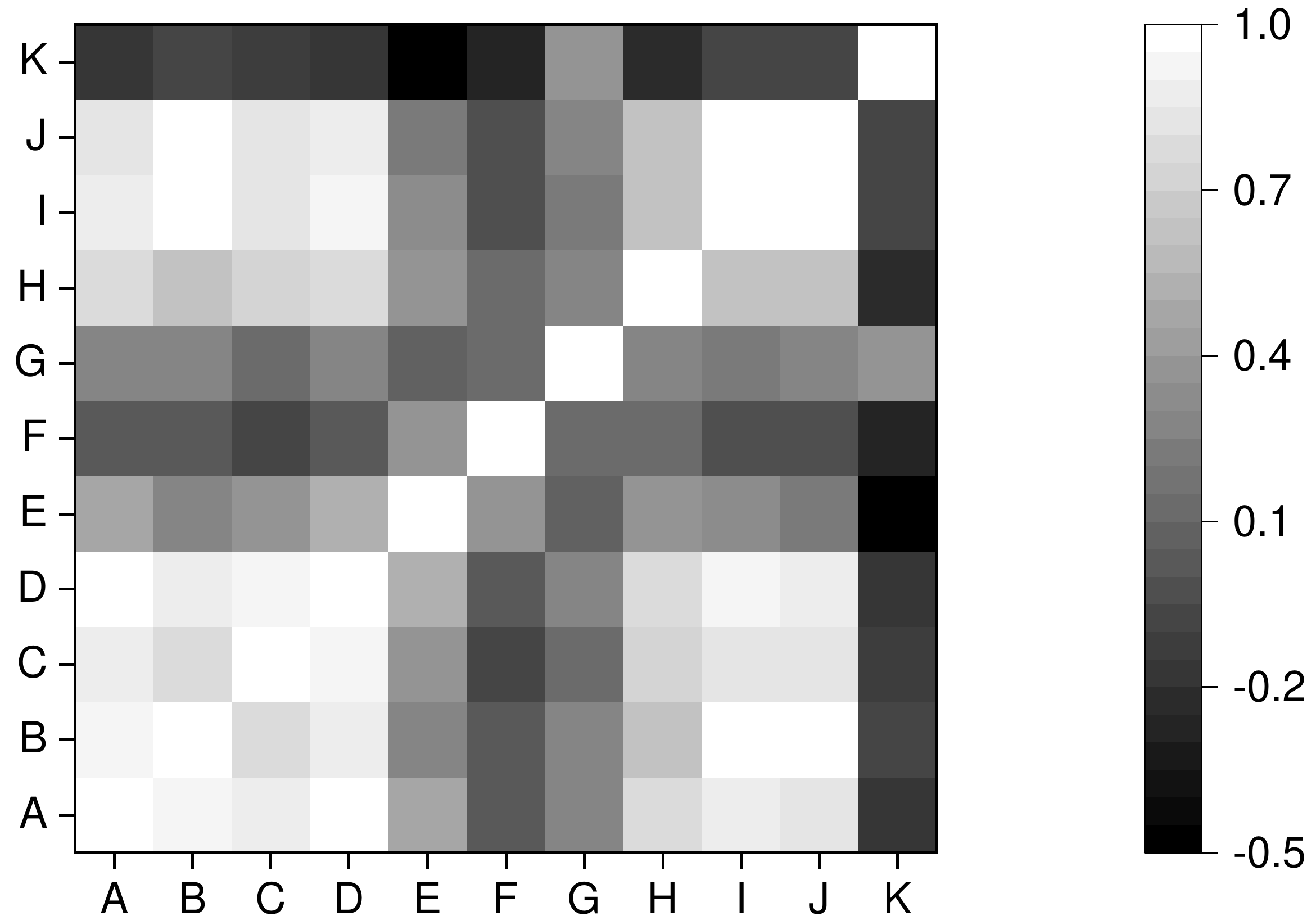}
    \caption{Correlation plot of input features (see legend in \emph{Table \ref{tab:legendcorrelation}}).}
    \label{fig:Correlation plot}
    \end{figure}
    
    \begin{table}[h]
\centering
\caption{\emph{Legend for \emph{Fig. \ref{fig:Correlation plot}}.} \label{tab:legendcorrelation}}
\scalebox{1}{
\begin{tabular}{l|l}
\hline
\textbf{Feature} & \textbf{Symbol}\\
\hline
Host Total energy (J) & A\\
Host EDP & B\\
Host Total DRAM access (GB) & C\\
Host FLOPs & D\\
Host GFLOP/s & E\\
Host FLOP/B & F\\
Ramulator IPC & G\\
Ramulator Total Time (ns) & H\\
Ramulator/DRAMPower Total trace energy (pJ) & I\\
Ramulator EDP & J\\
Speedup & K\\
   
\hline
\end{tabular}
}

\end{table}

In \emph{Fig. \ref{fig:Correlation plot}}, we show the correlation of the main features we used in this work. It may be easily visible that the correlation is equal to 1 for the same metric, while in the other cases is lower. Ramulator IPC is a key factor for making offloading decisions, and indeed it has the highest correlation with the EDP speedup. Since it is time-consuming to run Ramulator each time for a new unseen application or application with a different data set, we deploy an RF regression model to predict the Ramulator IPC and consequently predict the NMC suitability classification. This step is quicker than NMC simulation and enables early design exploration of unseen applications.

\subsubsection{NMC suitability prediction}
After the model is trained, validated and tuned, the final step is to test it on an unseen application. Similarly to \cite{8806888}, we trained the model using the data of all the application \emph{excluding} the one the model will predict. In this manner, the prediction will be more complex, and the application can be considered \emph{unseen}. 
Since Ramulator is time-consuming and, in particular, it takes several days to simulate for the radio-astronomical imaging algorithms, even with a small image such as 128x128 pixels, we used only Polybench applications for the training (excluding the predicted application if necessary). In particular, for this small dataset, more than 8640 minutes are necessary and large disk space is required, such as a few terabytes. Furthermore, we simulate the above-mentioned small dataset for Gridder and Degridder, which are well-known compute-bound application \cite{VEENBOER2020100386} and will not benefit from NMC in any case to prove the unsuitability of these kernel for NMC offloading. Indeed, their EDP speedup is small (near 0).

     \begin{figure}[h]
    \centering
    \includegraphics[width=6.5cm]{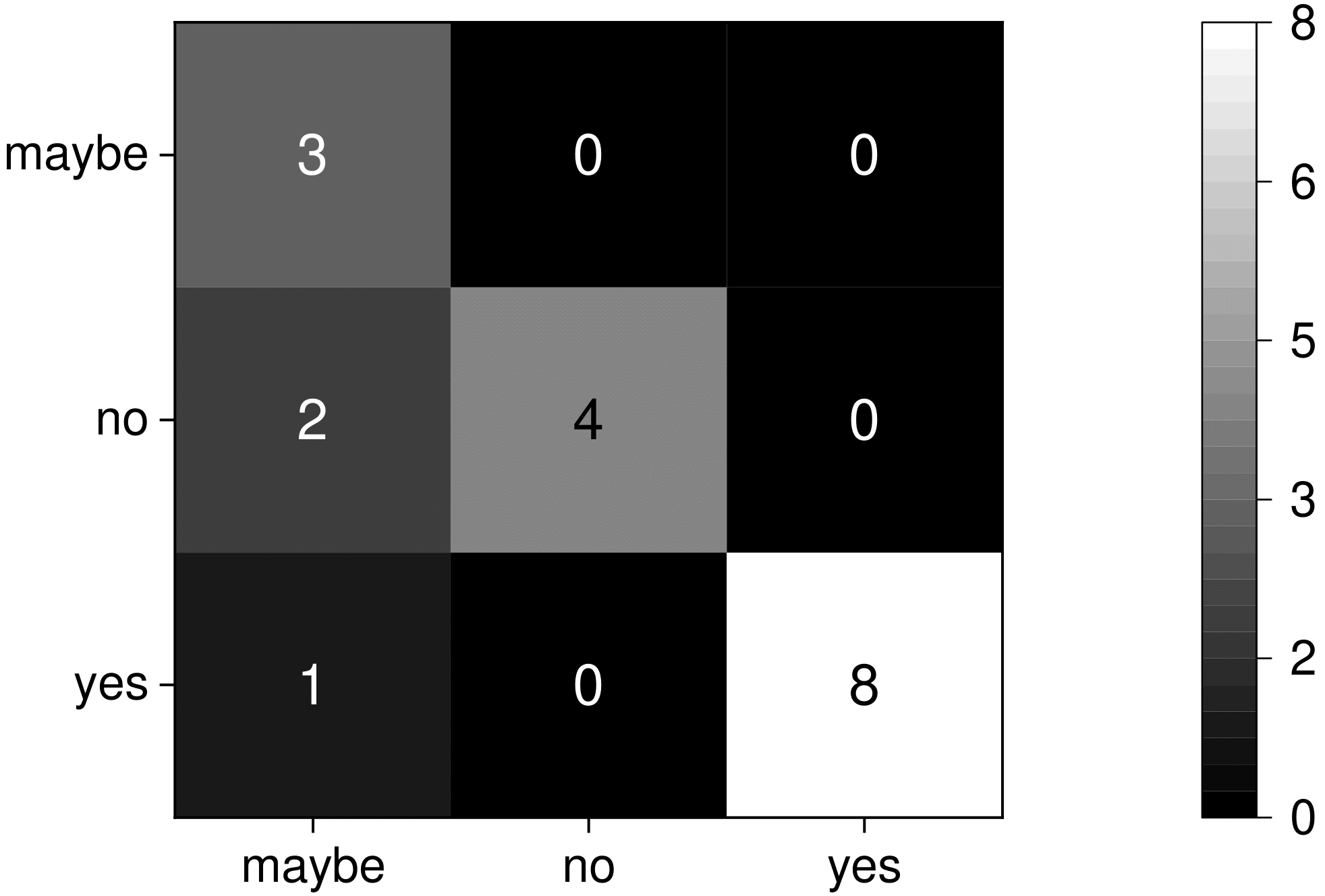}
    \caption{Confusion Matrix.}
    \label{fig:Confusion matrix}
    \end{figure}
    
The confusion matrix \emph{Fig. \ref{fig:Confusion matrix}} reports the distribution of true and false positive of the prediction done by NMPO. For instance, in the bottom row 9 tests are predicted correctly 8 times as ``yes" and 1 time as false positive ``maybe".
The prediction results are slightly related to the roofline model, which is still a good tool for application characterization. Indeed, compute-bound applications do not benefit from NMC, L3 memory bounded application benefit from NMC, and the other applications can benefit from NMC based on the dataset size as applications tend to incur frequent cache misses in L3 and stall on data to be fetched from DRAM~\cite{awan2015data}

  \begin{figure}[h]
\centering
\begin{subfigure}[b]{0.55\textwidth}
   \includegraphics[width=9cm]{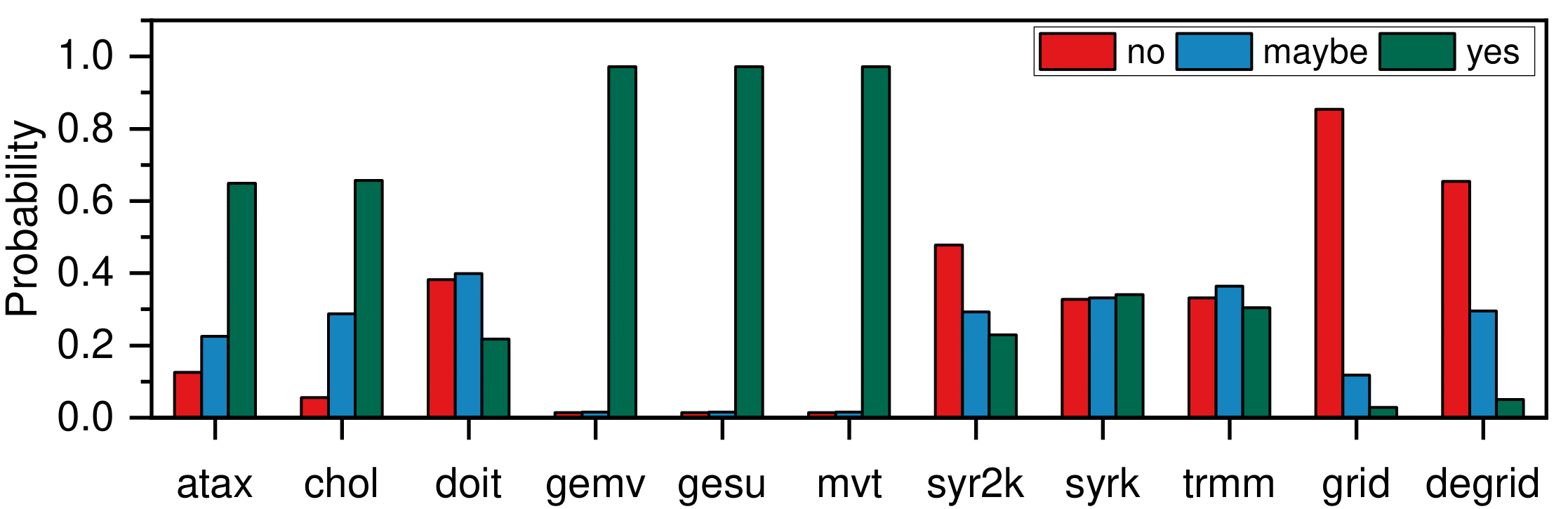}
   \caption{}
   \label{fig:threads8} 
\end{subfigure}

\begin{subfigure}[b]{0.55\textwidth}
   \includegraphics[width=9cm]{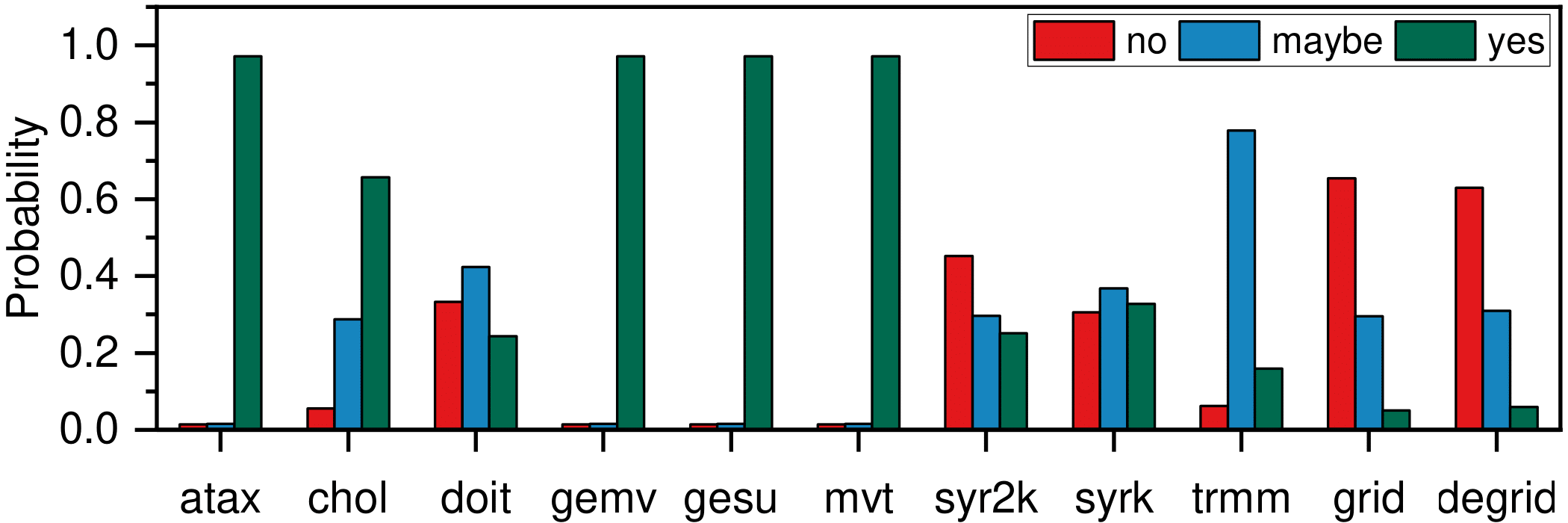}
   \caption{}
   \label{fig:threads16}
\end{subfigure}

\caption{Model probability of predictions: (a) 8 threads, and (b) 16 threads.}
\label{fig:probability}
\end{figure}

The machine learning model classification probability for the test cases is reported in \emph{Fig. \ref{fig:probability}} for both 8 and 16 threads test cases. For instance, for the atax test case using 8 threads, the probability of predicting ``yes" is about 60\%, while for mvt it is 100\%. This heavily depends on the training datasets employed.

\begin{figure}[h]
    \centering
    \includegraphics[width=9cm]{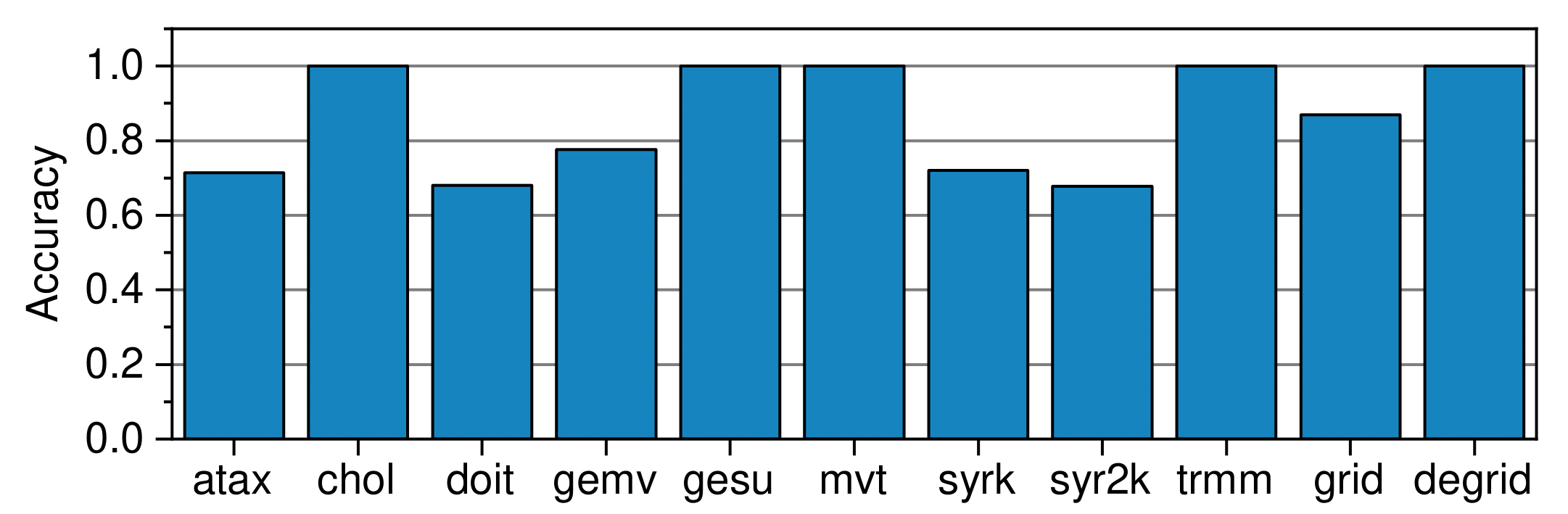}
    \caption{Accuracy of offloading using NMPO.}
    \label{fig:Accuracy of offloading using NMPO.}
    \end{figure}
    
The overall model accuracy is reported in \emph{Fig. \ref{fig:Accuracy of offloading using NMPO.}} per applications. While some applications have a 100\% accuracy, some of them are below 80\%. In average, the accuracy is 85.6\%.

\subsubsection{Improved estimation for training time}

\begin{figure}[h]
    \centering
    \includegraphics[width=9cm]{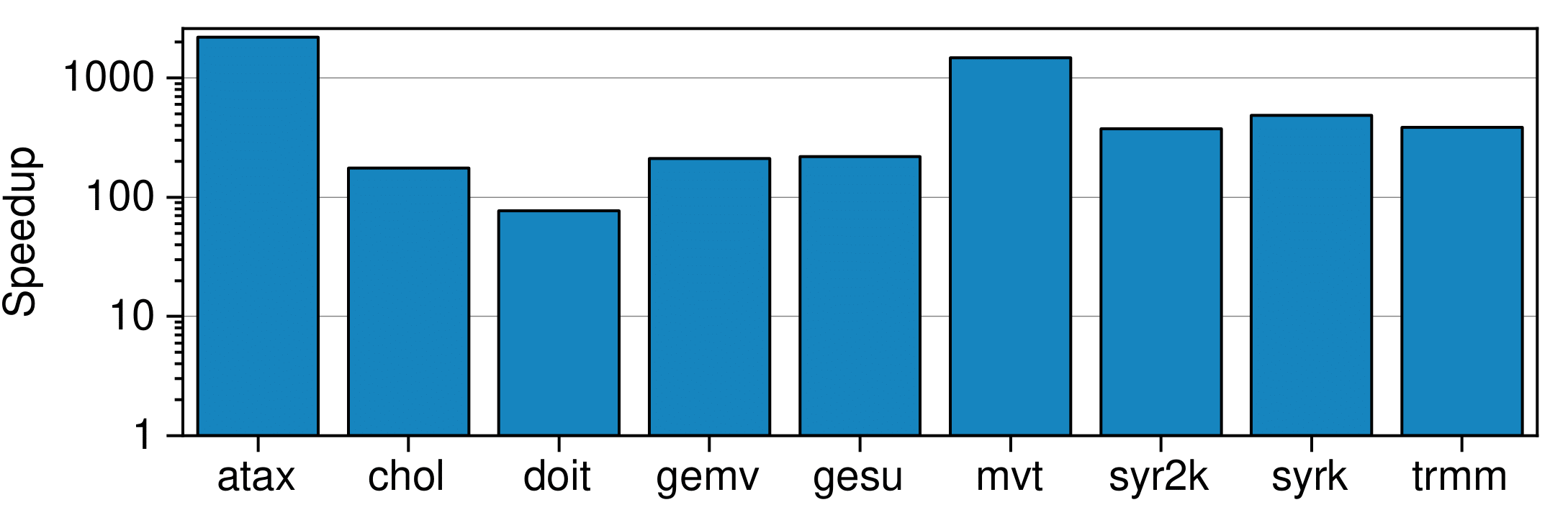}
   \caption{\emph{Perf vs PISA execution time comparison.}
    \label{fig:perfvspisaspeedup}}
\end{figure}

Similar to \cite{8806888} the main bottleneck in these design space explorations is usually located in the training phase, where the NMC system must be simulated. This procedure usually can take days for a single application for real-world datasets. 
Furthermore, in \cite{8806888} the prediction phases consist in characterizing the application employing PISA \cite{10.1145/2742854.2742859}. However, PISA is slower than PMUs and for specific applications needs more than \SI{64}{GB} of DDR4, making this step really challenging. We reported in \emph{Fig. \ref{fig:perfvspisaspeedup}} the execution time speedup of perf compared to PISA. We employed the datasets reported in \emph{Table \ref{tab:perfvspisadataset}}, which are smaller compared to the ones in \emph{Table \ref{tab:application_and_parameters}} but that has value since the PISA overhead increases with the dataset size. We can notice 2 to 3 order of magnitude improvement comparing perf to PISA, thus making the use of perf for the prediction phase more convenient.

\begin{table}[H]
\centering
\caption{\emph{Dataset employed for comparing perf and PISA.} \label{tab:perfvspisadataset}}
\scalebox{1}{
\begin{tabular}{c|c|c|c}
\hline
\textbf{Application} & \textbf{Dataset} & \textbf{perf time [s]} & \textbf{PISA time [s]}\\
\hline
atax & 2000 & 0.23 & 503.85\\
chol & 512 & 0.16 & 28.01\\
doit & 64 & 0.27 & 20.78\\
gemv & 2000 & 0.26 & 55.62\\
gesu & 2000 & 0.32 & 69.27\\
mvt & 2000 & 0.24 & 356.16\\
syrk & 512 & 0.54 & 201.25\\
syr2k & 512 & 0.86 & 416.44\\
trmm & 512 & 0.36 & 140.6\\
\hline
\end{tabular}
}

\end{table}

\section{Related Work}
\label{sec:relatedwork}

Near-memory computing past works focused mainly on selecting specific memory-bound applications and optimize them with custom architectures on the logic-layer of 3d-stacked memory \cite{SINGH2019102868}. A few of them focused on offloading mechanisms or performance prediction to decide if the NMC system's scheduling is beneficial. We summarize the main related work on application offloading on NMC systems in \emph{Table \ref{tab:relatedwork}}. 

\begin{table}[h]
\centering
\caption{\emph{NMC offloading related work.} \label{tab:relatedwork}}
\scalebox{0.85}{
\begin{tabular}{l|l|l|l|l}
\hline
\textbf{Name} & \textbf{Year} & \textbf{Offloading} & \textbf{Accelerator} & \textbf{Memory}\\
\hline
Zhang et al. \cite{10.1145/2600212.2600213} & 2014 & estimation model & GPU & HMC\\
Ahn et al \cite{7551394}  & 2015 & compiler  and run-time & GPU & HMC \\
Hsieh et al. \cite{7284077} & 2016 & run-time & Ooo cores & HMC\\
Hadidi et al. \cite{10.1145/3155287} & 2017 & compiler & Fixed function units & HMC\\
Ahmed et al. \cite{8714956} & 2019 & compiler &  Fixed function units & HMC\\
Corda et al. \cite{8875040} & 2019 & PCA & in-order cores & HMC\\
Singh et al. \cite{8806888} & 2019 & ML model & in-order cores & HMC\\
\hline
\end{tabular}
}

\end{table}

Zhang et al. \cite{10.1145/2600212.2600213} employ a performance prediction model to decide how to schedule applications on their GPU-based NMC architecture.
Ahn et al. \cite{7551394} propose an offloading ad data mapping mechanism hidden to the programmer. This compiler-based mechanism can efficiently schedule workloads on their NMC-GPU system employing metrics such as memory bandwidth cost-benefit and memory mapping benefits.
Hsieh et al. \cite{7284077} propose an ISA extension to support NMC execution on an NMC system consisting of Ooo cores and HMC. The programmer must use this specific instruction to offload specific instruction to the NMC architecture.
Hadidi et al. \cite{10.1145/3155287} extend GraphPIM \cite{7920847} propose a compiler-based mechanism for instruction offloading on CPU/GPU-NMC systems.
Ahmed et al. \cite{8714956} propose a compiler-based mechanism able to detect code sections that reduce the off-chip data movement when accelerated on a CPU connected to HMC.
Corda et al. \cite{8875040} employ PISA-NMC \cite{10.1145/3323439.3323988}, an extended version of PISA capable of extracting metrics related to memory and task parallelism, to evaluate the correlation of these metrics and the NMC offloading suitability using the Principal Component Analysis (PCA).
Singh et al. \cite{8806888} design a high-level framework for predicting unseen application performance on an NMC system. This framework consists of a tuned random-forest model trained with hardware-independent feature and performance on an NMC system with HMC and in-order cores. While the model is capable of predicting the energy-delay-product accurately, prediction is slow. Indeed, this prediction needs to gather the hardware-independent feature of the unseen application using PISA \cite{10.1145/2742854.2742859}, which may take from 2 to 3 orders of magnitude compared to the application's execution time in the host system as we show in \emph{Section \ref{subsec:offloading}}. We use the hardware-dependent application features collected with a small execution time overhead to predict the NMC offloading suitability to overcome this critical issue. Furthermore, while in \cite{8806888} specific datasets are so small that they cannot be sampled by perf the PMUs (execution time lower than 0.001s), we selected large datasets that can generate DRAM traffic. This makes it possible to evaluate which applications are suitable for NMC offloading when accessing external DRAM.

Performance prediction of unseen applications on specific architectures is a widely researched topic. However, just some of them focus on NMC. Indeed, as shown in \emph{Table \ref{tab:relatedwork2}}, most of them focus on CPU and GPU as target offloading architecture. Concerning the machine learning model employed, in past work, linear regression, ANN and random forest have been employed with different tuning options. Similar to Singh et al. \cite{8806888} and Mariani et al. \cite{7973739} we employ the random forest algorithm because it can achieve higher prediction accuracy.

\begin{table}[h]
\centering
\caption{\emph{Performance prediction related work.} \label{tab:relatedwork2}}
\scalebox{1}{
\begin{tabular}{l|l|l|l}
\hline
\textbf{Name} & \textbf{Year} & \textbf{ML model} & \textbf{Architecture}\\
\hline
Joseph et al. \cite{1598116} & 2006 & Linear Regression & CPU\\
Calotoiu et al \cite{6877478} & 2013 & Empirical model & CPU \\
Bailey et al. \cite{6957246} & 2014 & Linear Regression & CPU/GPU \\
Wu et al. \cite{7056063} & 2015 & ANN & GPU\\
Mariani et al. \cite{7973739} & 2017  & Random Forest & Cloud HPC\\
Singh et al. \cite{8806888} & 2019 & Random Forest & NMC\\
\hline
\end{tabular}
}

\end{table}
\section{Conclusion}
\label{sec:conclusion}
We present NMPO, a high-level framework based on ensemble learning models and hardware-dependent profiling that facilitate quick and precise predictions to offload suitable applications to NMC kernels. This framework aids in the early design stage exploration of unseen applications on modern DRAMs like HMC.
Unlike slow simulators, NMPO employs an ensemble learning technique called Random Forest with hyper tuning to speculate the offloading of an application. Furthermore, NMPO is much faster than the current state-of-the-art NMC simulator, and other machine learning-based frameworks with platform-independent profiling since hardware-dependent characterization used in NMPO has far less execution time overhead than hardware-independent ones. Thus, NMPO with 85.6\% accuracy, quicker analysis and user-friendliness is the go-to ML-based framework for early design stage exploration.

\section*{Acknowledgments}
This work is funded by the European Commission under Marie Sklodowska-Curie Innovative Training Networks European Industrial Doctorate (Project ID: 676240). We would like to thank Gabor Nemeth from Ericsson Research for his feedback on the draft of the paper.

\bibliographystyle{IEEEtran}
\bibliography{IEEEabrv,refshort}

\end{document}